\documentclass[12pt,preprint]{aastex}
\usepackage{underscore}
\usepackage{color}

\slugcomment{Accepted on December 4, 2015}

\shorttitle{17P/Holmes Outburst Remnant Dust}
\shortauthors{Ishiguro et al.}

\begin{document}

\title{Detection of Remnant Dust Cloud Associated with the 2007 Outburst of 17P/Holmes}

\author{Masateru \textsc{Ishiguro}}
\affil{Department of Physics and Astronomy, Seoul National University,
Gwanak, Seoul 151-742, South Korea}

\author{Yuki \textsc{Sarugaku}}
\affil{Kiso Observatory, Institute of Astronomy, Graduate School of Science,
The University of Tokyo, Mitake, Kiso-machi, Nagano, Nagano 397-0101,
Japan}

\author{Daisuke \textsc{Kuroda}}
\affil{Okayama Astrophysical Observatory, National Astronomical
Observatory of Japan, Asakuchi, Okayama 719-0232, Japan}

\author{Hidekazu \textsc{Hanayama}}
\affil{Ishigakijima Astronomical Observatory, National Astronomical
Observatory of Japan, 1024-1 Arakawa, Ishigaki, Okinawa 907-0024, Japan}

\author{Yoonyoung \textsc{Kim}, Yuna \textsc{Kwon}}
\affil{Department of Physics and Astronomy, Seoul National University,
Gwanak, Seoul 151-742, South Korea}

\author{Hiroyuki \textsc{Maehara}}
\affil{Okayama Astrophysical Observatory, National Astronomical
Observatory of Japan, Asakuchi, Okayama 719-0232, Japan}

\author{Jun \textsc{Takahashi}}
\affil{Nishi-Harima Astronomical Observatory, Center for Astronomy,
University of Hyogo, Sayo, Hyogo 679-5313, Japan}

\author{Tsuyoshi \textsc{Terai}}
\affil{Subaru Telescope, National Astronomical Observatory of Japan,
Hilo, HI 96720, USA}

\author{Fumihiko \textsc{Usui}}
\affil{Department of Astronomy, Graduate School of Science,
The University of Tokyo,\\ 7-3-1 Hongo, Bunkyo-ku, Tokyo 113-0033, Japan}

\author{Jeremie J. \textsc{Vaubaillon}}
\affil{Observatoire de Paris, I.M.C.C.E., Denfert Rochereau, Bat. A.,
FR--75014 Paris, France}

\author{Tomoki \textsc{Morokuma}, Naoto \textsc{Kobayashi}}
\affil{Institute of Astronomy, Graduate School of Science, The University of Tokyo,
2-21-1 Osawa, Mitaka, Tokyo 181-0015, Japan}

\author{Jun-ichi \textsc{Watanabe}}
\affil{National Astronomical Observatory of Japan, Mitaka, Tokyo
181-8588, Japan}

\begin{abstract}
This paper reports a new optical observation of 17P/Holmes one orbital period after the historical
outburst event in 2007. We detected not only a common dust tail near the nucleus, but also a long narrow
structure that extended along the position angle 274.6\arcdeg $\pm$ 0.1\arcdeg\
beyond the field of view of the Kiso Wide Field Camera, i.e., $>$0.2\arcdeg\
eastward and $>$2.0\arcdeg\ westward from the nuclear position. The width of the structure decreased
westward with increasing distance from the nucleus. We obtained the total cross
section of the long extended structure in the field of view, $C_\mathrm{FOV}$ = (2.3 $\pm$ 0.5) $\times$ 10$^{10}$ m$^2$.
From the position angle, morphology and the mass, we concluded that the long narrow structure consists of materials ejected during the 2007
outburst. 
On the basis of the dynamical behavior of dust grains in the solar radiation field, we estimated that the long narrow structure would be composed of 1 mm--1 cm
grains having an ejection velocity of $>$50 m s$^{-1}$. The velocity was more than one order of magnitude faster
than that of millimeter -- centimeter grains from typical comets around a heliocentric distance $r_\mathrm{h}$ of 2.5 AU.
We considered that sudden sublimation of a large amount of water ice ($\approx$10$^{30}$ mol s$^{-1}$) would be responsible for
the high ejection velocity. We finally estimated a total mass of $M_\mathrm{TOT}$=(4--8) $\times$ 10$^{11}$ kg and a total kinetic energy of $E_\mathrm{TOT}$=(1--6)$\times$10$^{15}$ J for the 2007 outburst ejecta,
which are consistent with those of previous studies that conducted soon after the outburst.
\end{abstract}
\keywords{comets: individual (17P/Holmes) --- interplanetary medium --- meteorites, meteors, meteoroids}


\section{INTRODUCTION}
\label{sec:introduction}

This paper provides a new optical observation of 17P/Holmes during the 2014 perihelion passage using
a wide-field camera recently available at the Kiso Observatory, when it was located at the position of the 2007 outburst. 

An unprecedented cometary outburst occurred at 17P/Holmes on UT 2007 October 23,
brightening by about one million times within a day \citep{Sekanina2009,Hsieh2010}.
Soon after, the comet was enclosed by an envelope composed of high-speed dust grains.
The envelope faded out in about a year because of solar radiation pressure as well as its high ejection velocity,
leaving behind a near-nuclear dust cloud. Afterward, the comet remained active for years, showing
a minor outburst \citep{Stevenson2010} and lingering dust ejection at 4--5 AU \citep{Ishiguro2013}.
An analysis of the faint dust tail in the infrared suggested that it could contain trailing dust particles ejected during
the 2007 outburst \citep{Stevenson2014}.

Big particles from comets are widely observed through telescopic observations and remote-sensing observations with spacecrafts
\citep[see e.g.,][]{Ishiguro2002,Rotundi2015}. These large particles can stay close to the orbits of parent bodies for many
revolutions around the Sun and form `dust trails', which are occasionally discriminated from 'dust tails' consisted of fresh dust particles
ejected during  current returns. Neck-line is a sub-structure rarely detected in dusts tails (and in dust trails, in principle), which is caused by a dynamical effect. Theoretically, dust particles ejected at a point (first node) converge on the orbital plane
of the parent body at the opposite end viewed from the Sun (the second node, i.e. the differential true anomaly
between the dust ejection point and observed location is 180\arcdeg). The accumulation of the dust
particles enhances the surface brightness of the dust cloud, as first proposed by
\citet{Kimura1977} \citep[see also,][]{Fulle1988}. An analogous phenomenon essentially occurs when dust particles complete one orbital revolution (i.e. the third node).

Taking advantage of the neck-line effect, we aimed to detect the debris cloud ejected the 2007 outburst.
We described the observation and data reduction in Section 2,  diagnosed the observed image in Section 3 on the basis
of photometric and dynamical properties. As the result, we could obtain indubitable evidence for the debris cloud
associated with the 2007 outburst. We derived the velocity, mass and total kinetic energy of the outburst grains
using our new observation data to deepen our understanding of the historical event.

\section{OBSERVATION AND DATA ANALYSIS}
\label{sec:observation}

\begin{figure}
 \epsscale{1.0}
   \plotone{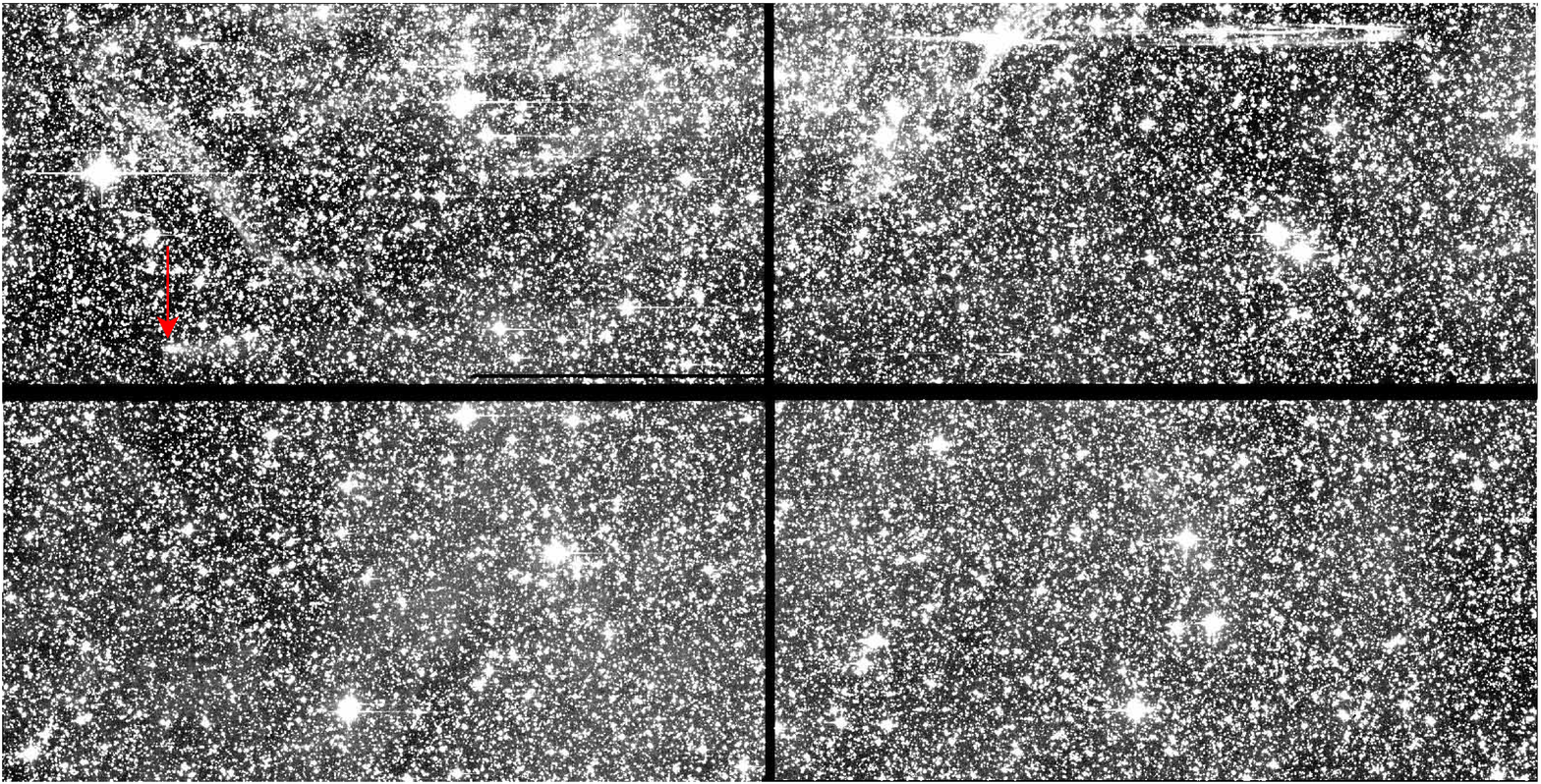}
  \caption{Example of an $R_\mathrm{C}$-band (wavelength 0.64 \micron) full-scale (2.2\arcdeg $\times$ 1.1\arcdeg)
  snapshot image taken at UT 17:09, 2014 September 22. Location of the comet is indicated by arrow.
  Cometary tail was unclear in the snapshot because of contamination by stars and diffuse galactic light.}
  \label{fig:f1}
\end{figure}

We imaged 17P/Holmes in the $R_\mathrm{C}$ band for four nights on UT 2014 September 18, 22--23, and 25
with the Kiso Wide Field Camera (KWFC) attached to the 105 cm Schmidt telescope operated by the Kiso Observatory,
the University of Tokyo. The KWFC is a mosaic CCD camera consisting of eight CCD chips with a total of 8k $\times$ 8k
pixels \citep{Sako2012}. The combined system provides a 2.2\arcdeg $\times$ 2.2\arcdeg\ field of view (FOV) with
a moderate pixel resolution (0.946\arcsec/pixel). We used the lower half of the KWFC for our observation
to reduce the readout time (from 144 to 68 s), so the instrument covered 2.2\arcdeg $\times$ 1.1\arcdeg\ on the sky plane
in a single snapshot with four CCD chips (see Figure \ref{fig:f1}). Because the telescope can be operated only in a sidereal tracking mode, we could not take longer exposures. We set the individual exposure time to 180 s, during which time
the comet moved by 1.5\arcsec, which is larger than the pixel scale but still less than the half of the typical seeing disk size
at the observatory \citep[i.e., a full width at half-maximum (FWHM) of 3.3--5.3\arcsec,][]{Morokuma2014}. We took images of 17P/Holmes on UT 2014
September 18 and 22--23 in the range between $\sim$0.2\arcdeg\  eastward and $\sim$2\arcdeg\ westward from the nucleus
position. In addition, we contrived to take images on UT 2014 September 25 with a wide coverage, within 6\arcdeg\ westward
from the nucleus, scanning along the projected orbit of the comet.

Only the images on UT 2014 September 22 are available for study  because the sky background was too high
to detect the faint structure on UT 2014 September 18 owing to moonlight, the weather conditions were bad because of a typhoon
(known as Tropical Storm Fung-wong) on UT September 23, and the total exposure time was insufficient to subtract the background
objects on UT 2014 September 25. For these reasons, we focused on analysis of images taken on UT 2014 September 22. On that night, we took
60 snapshot images (i.e., a total exposure time of 3 h) at UT 15:11--19:35, when the comet was at a heliocentric distance
$r_\mathrm{h}$ of 2.464 AU, an observer's distance $\Delta$ of 2.094 AU, and a phase angle (Sun--comet--observer angle) $\alpha$ of 23.7\arcdeg.
The true anomaly was $\theta_\mathrm{TA}$ = 63.1\arcdeg, slightly larger than that of the 2007 outburst, $\theta_\mathrm{TA}$ = 61.1\arcdeg.
In this configuration, the convergent point exists at 2.7\arcdeg\  westward from the nucleus.

The observed data were preprocessed using bias and dome flat images. Figure \ref{fig:f1} shows an example snapshot image
after bias and flat field correction. There are $\sim$1\arcmin\ gaps between each CCD chip. Because we employed the dithering
operation mode for the telescope, the gaps are mostly (but not perfectly) eliminated by multiple exposures. In the snapshot,
the comet and dust cloud are unclear because they were located in the
star-crowded area near the galactic plane (galactic latitude $b\sim2$\arcdeg). Accordingly, we paid close attention to
elimination of background components. We thus subtracted the background stars and diffuse interstellar cirri using images
taken on the next night (UT 2014 September 23) when the comet had moved northeastward by 15\arcmin.  Moreover, we masked
the position of stars brighter than R$_\mathrm{C} \sim$ 22 in the images after star and cirrus subtraction to eliminate the remnants
caused by misalignment of the stellar positions. We then combined 60 masked images according to the motion of the comet nucleus, excluding
the masked regions. The defective pixels and CCD gaps are also excluded by the image combination process.
Flux calibration was conducted using field stars archived in the UCAC3 catalog, ensuring a photometric
accuracy of $\sim$0.1 mag or less \citep{Zacharias2010}. Geometrical correction was performed by comparison with the astrometric
data in the USNO B1.0 catalog \citep{Monet2003}.  The positional error of the USNO B1.0 catalog, $\sim$0.4\arcsec, is good enough
that we can discuss the position angle and morphology of the dust structure in this study.
\label{sec:observations}

\section{RESULTS}
\label{sec:results}

\subsection{Appearance}
\label{subsec:appearance}
\begin{figure}
\begin{center}
 \epsscale{1.0}
  \plotone{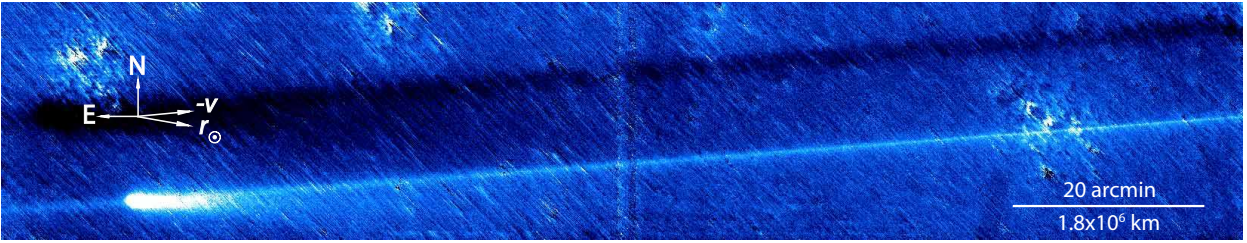}
  \caption{Processed $R_\mathrm{C}$-band image of 17P/Holmes taken
  on UT 2014 September 22  when the comet was at $r_\mathrm{h}$=2.464 AU, $\Delta$=2.094 AU, and $\alpha$=23.7\arcdeg.
   The shadowed region is caused by image subtraction using an image
  taken on UT 2014 September 23 (see Section 2). Hatched diagonal lines from upper left to lower right
  are remnants of background stars which could not be subtracted using our data reduction algorithm.
  Vertical lines in the center are a relic of the CCD gap. Antisolar direction and negative velocity
  vector are indicated by ``$r_\odot$'' and ``$-v$,'' respectively.
  The image has a standard orientation in the sky; that is, north is up, and east is to the left.
  The FOV is 130\arcmin $\times$ 25\arcmin.   }
    \label{fig:f2}
    \end{center}
\end{figure}

Figure \ref{fig:f2} shows a composite image taken on UT 2014 September 22 after background objects and
instrumental artifacts were subtracted. The figure shows not only a cometary tail near the nuclear position (a whitish cloud in the lower left
corner) but also a long structure extending from the lower left (southeast) to the upper right (northwest). It spreads to both sides
beyond the FOV of the KWFC ($>$2.2\arcdeg). We hereafter analyze these cloud
morphologies as shown below.

\subsubsection{Near-Nuclear Dust Tail and Coma}
\label{subsubsection:tail}
\begin{figure}
 \epsscale{1.0}
   \plotone{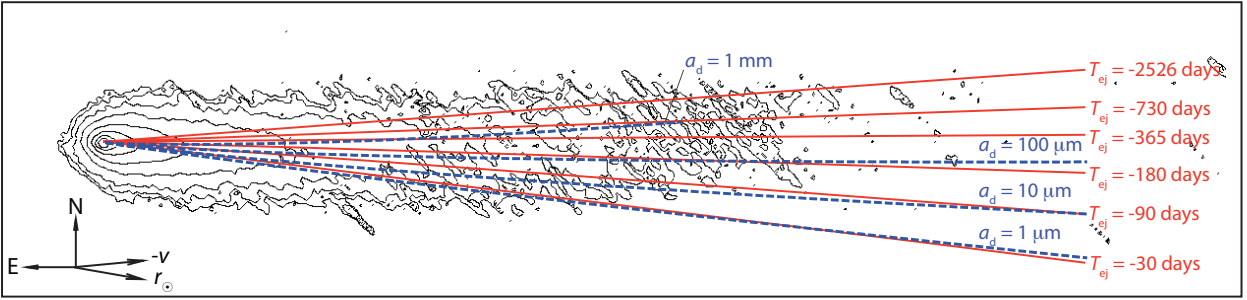}
  \caption{Synchrones and syndynes with contour map of the near-nuclear dust tail. Numbers for synchrones denote the days of dust ejection
  before the time of observation. $T_\mathrm{ej}$ = $-2526$ days corresponds to the day of the 2007 outburst. We assumed the mass density of the dust particles  $\rho_\mathrm{d}$ = 10$^3$ kg m$^{-3}$   when the $\beta$ values were converted to the radius $a$.   The FOV of the contour
  is 20\arcmin $\times$ 5\arcmin\ (3\% of the area of Figure \ref{fig:f2}).
  Note that the long extended structure does not appear in the contour map  because we showed the bright part of the dust tail.}
    \label{fig:f3}
\end{figure}

Figure \ref{fig:f3} shows a close-up of the contours of the near-nuclear dust tail.
It extended between the negative heliocentric velocity
vector (at a position angle, P.A.=275.4\arcdeg) and the antisolar direction (P.A. = 260.1\arcdeg). In Figure \ref{fig:f3}, we
show sets of synchrones (loci of positions of particles having a wide range of sizes released at given times, $T_\mathrm{ej}$) and syndynes (loci of
positions of particles of a given size, $a_\mathrm{d}$, having a wide range of ejection epochs). Although these lines are not separated well, we roughly
estimated the dust ejection epoch and particle size. We determined the locus of the maximum brightness of the dust tail using least square fitting
with a Gaussian function in every 1\arcmin\ bin and found that the dust tail extended to P.A. =270\arcdeg$\pm$0.5\arcdeg,
which corresponds to the synchrone of $T_\mathrm{ej}\sim180-360$ days before the observation. Because the comet was observed
179 days after the perihelion passage, it is likely that the near-nuclear dust tail consisted of particles ejected during the perihelion
passage in 2015. From comparison with the syndynes, we estimate an effective particle radius of 10 \micron\--1 mm for the near-nuclear tail.

We visually set the sunward extent of the dust coma to $\sim$40\arcsec, which corresponds to the distance projected on the sky plane,
$l$ = 6 $\times$ 10${^7}$ m, at the position of the comet. Considering $l$ as the turnaround distance of dust particles ejected toward
the solar direction while being pushed by solar radiation pressure, we placed a limitation on the ejection velocity,
$V$ = 340$\sqrt{\beta}$ m s$^{-1}$, where $\beta$ denotes the ratio of the solar radiation pressure with respect to the solar gravity \citep{Jewitt1987}.
In the possible size range (10 \micron\--1 mm, or $\beta$ = 5.7 $\times$ 10$^{-2}$--5.7 $\times$ 10$^{-4}$), we estimated
an ejection velocity $V$ of 8 m s$^{-1}$ for 1 mm particles and 80 m s$^{-1}$ for 10 \micron\ particles.
Although we understand that these are very crude estimates,
the derived velocity is typical of dust emission from Jupiter-family comets (JFCs) at $r_\mathrm{h}\sim$ 2.5 AU \citep[][]{Ishiguro2007, Ishiguro2008}.
Thus, 17P/Holmes changed from its peculiar appearance just after the 2007 outburst to an appearance typical of JFCs
in only one orbital revolution.

We measured $Af\rho$ values with differential aperture size from
5,000 km (3.3\arcsec, equivalent to 1$\times$FWHM) to 50,000 km at
intervals of 5,000 km, and obtained almost constant values of $Af\rho$=139--141 cm
within 10,000 km but found significant drops due to the radiation pressure.
The $Af\rho$ value is typical of general comets listed in \citet{AHearn1995} ($\approx$10$^2$--10$^3$ cm
around 2.5 AU). 

\subsubsection{Long Extended Structure}
The prominent feature is the long narrow structure.
There could be four possibilities to create such long extended structure: (1) dust tail (i.e. dust particles ejected during
the current return), (2) ion tail, (3) dust trail, and (4) neck-line. We measured a position angle of 274.6\arcdeg $\pm$ 0.1\arcdeg,
which deviates significantly from
that of the antisolar direction (P.A. = 260.1\arcdeg) but is very close to that of the negative heliocentric velocity vector
(P.A. = 275.4\arcdeg). It also coincides with the direction of the synchrone of the 2007 outburst epoch (i.e., $T_\mathrm{ej}$ = $-$2526 days
in Figure \ref{fig:f3}), although the time resolution of the synchrone analysis is not sufficiently accurate to specify the exact ejection
epoch. This fact indicates that the long narrow structure
is not associated with neither (1) dust tail nor (2) ion tail, but with either (3) dust trail or (4) neck-line (i.e. a swarm of dust particles
ejected during the last perihelion passage or even before). It is interesting to think why the comet has this spectacular long extended structure,
although the near-nuclear dust tail looks typical of JFCs, as we mentioned above (Section \ref{subsubsection:tail}).

\begin{figure}
 \epsscale{0.8}
    \plotone{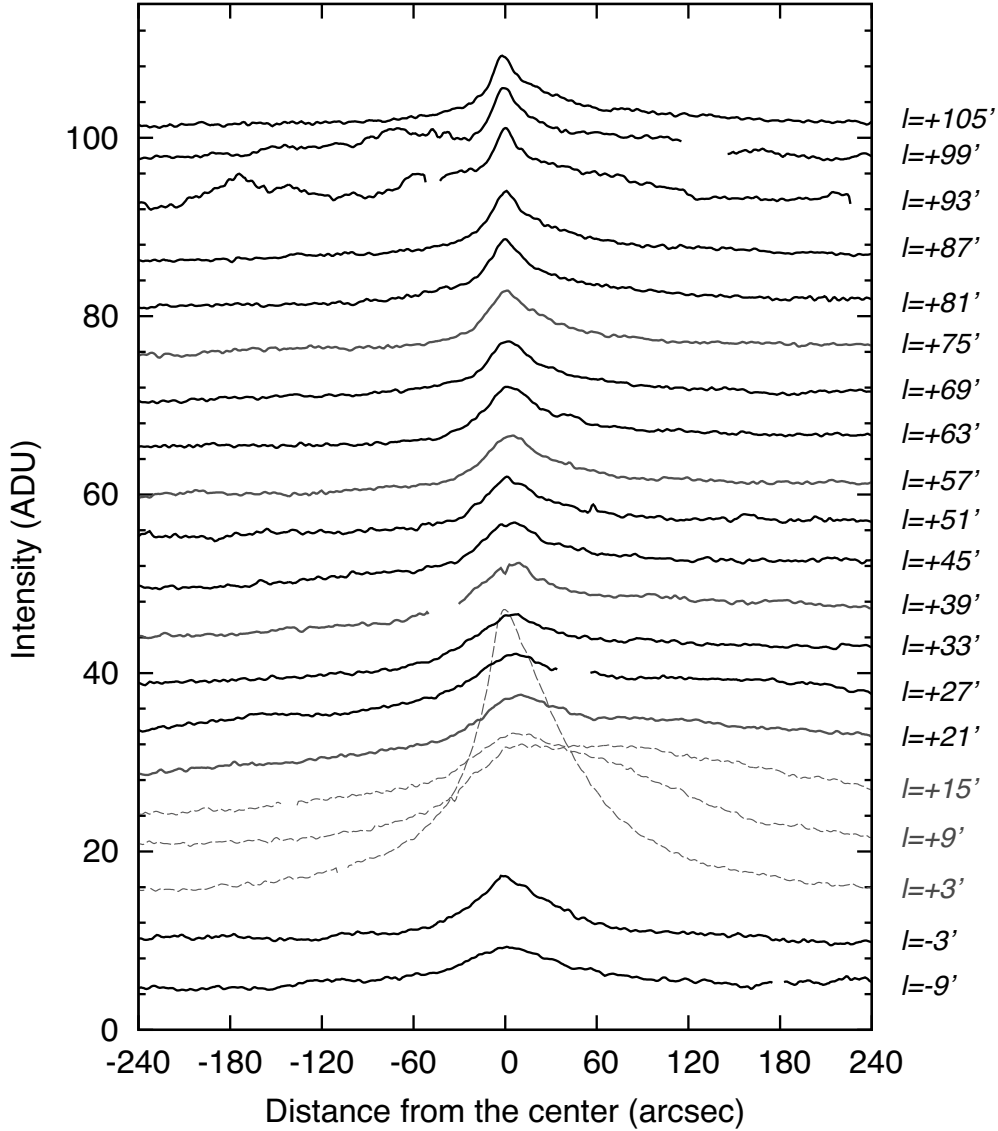}
  \caption{Cut profiles perpendicular to the long extended structure averaged along the extended direction of
  the structure within the bin length of 6\arcmin. Horizontal axis represents the distance from the central position of the long extended structure,
  where negative values are to the north and positive values are to the south.  $l$ denotes the distance from the position of the nucleus
  to the center of the cut profiles. Negative $l$ values are in the leading (leftward) direction, whereas positive $l$ values are in the
  trailing (rightward in Figure 1) direction of the orbital motion. For visibility, these profiles are offset by 5 ADU, where 1 ADU corresponds
  to 28.5 mag/arcsec$^{2}$, or 3.83 $\times$ 10$^{-9}$ W m$^{-2}$ sr$^{-1}$ \micron$^{-1}$, or 522 Jy sr$^{-1}$}.
    \label{fig:cut}
\end{figure}

\begin{figure}
 \epsscale{0.8}
   \plotone{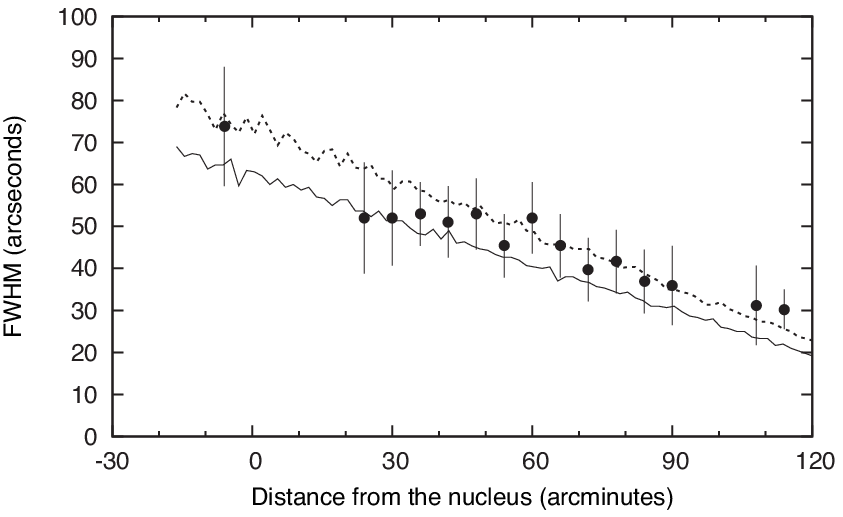}
   \plotone{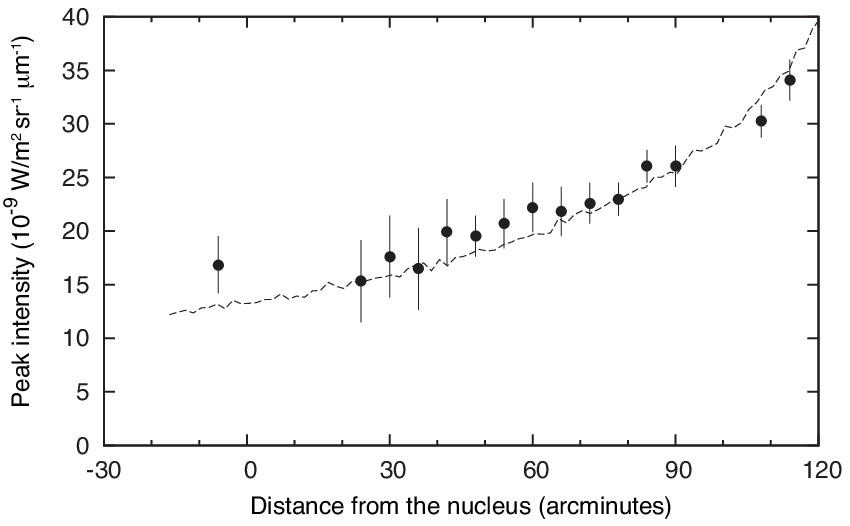}
  \caption{(top) FWHM and (bottom) peak brightness of the long extended structure with respect to the apparent distance from
  the 17P/Holmes nucleus. Positive and negative values of the distance denote the trailing and leading directions, respectively.
  Error bars represent the uncertainty associated with the sky background subtraction, as well as the statistical noise and calibration error.
  Lines are example results of the model fitting described in Section \ref{sec:model}. Solid  line: $V$ = 50 m s$^{-1}$ and $q$ = $-3.4$, dashed line: $V$ = 60 m s$^{-1}$ and $q$ = $-3.4$.}
    \label{fig:fwhm}
\end{figure}

We examined the surface brightness and width. Figure \ref{fig:cut} shows the surface
cut profiles perpendicular to the long extended structure, where we averaged the brightness along the extended direction  (P.A. = 274.6\arcdeg)
in the bin length of 6\arcmin.
The structure was unclear at the distance $+3$\arcmin$<l<+9$\arcmin\ because the bright near-nuclear dust tail overlapped
the faint extended structure. We fitted the background sky brightness by third-order polynomials and obtained the peak brightness and
FWHM. Figure \ref{fig:fwhm} shows the result. The FWHM clearly increased as the peak brightness
decreased from west to east. The peak brightness is in the range of 26.2--27.0 mag/arcsec$^2$,
which is equivalent to those of some bright cometary dust trails  such as 67P/Churyumov--Gerasimenko \citep{Ishiguro2008} and 22P/Kopff
\citep{Ishiguro2002}. It is, however, important to notice that the shape and brightness distributions differ from general dust trails. 
As shown in some papers  \citep[see e.g.,][]{Sykes1990,Ishiguro2002,Ishiguro2003}, dust trails show narrowing toward the nuclei.
In addition, the fading toward the nucleus  is inconsistent with the dust trail structure of 17P/Holmes detected by Spitzer observation, where the 
brightness increased toward the nucleus \citep{Reach2010}.
Thus, the long extended structure in our 17P/Holmes image could be unlike the typical dust trail structures seen to date.

\subsection{Photometry of the Long Extended Structure}
To determine the surface brightness profile of the long extended structure, we summed up the signal from the extended structure. We set a rectangular aperture box of
2\arcmin $\times$ 2.2\arcdeg. After subtracting the sky background, which was determined at 1--3\arcmin\ from the trail center on
both the north and south sides, we obtained the total $R_\mathrm{C}$ magnitude of $m_\mathrm{R}$ = 12.3 $\pm$ 0.2 in the  KWFC FOV. The observed
$R_\mathrm{C}$ magnitude was converted to the absolute magnitude (i.e., corrected to unit heliocentric and  geocentric distances
at zero phase angle) using

\begin{eqnarray}
H_\mathrm{R}=m_\mathrm{R} - 5~\log_{10}\left(r_\mathrm{h} \Delta\right)-2.5~\log_{10}\left(\Phi\left(\alpha\right)\right),
\label{eq:HR}
\end{eqnarray}

\noindent
where the term $\log_{10}\left(\Phi\left(\alpha\right)\right)$ characterizes the scattering phase function of dust grains.
Using a commonly used formula, $\log_{10}\left(\Phi\left(\alpha\right)\right)$ = 0.035$\alpha$, and substituting $r_\mathrm{h}$ = 2.464 AU,
$\Delta$ = 2.094 AU, and $\alpha$ = 23.7\arcdeg, we obtained $H_\mathrm{R}$ = $m_\mathrm{R} -$ 4.4 = 7.9 $\pm$ 0.2. The absolute magnitude is
converted into the cross section by the following equation \citep{Russell1916}:

\begin{eqnarray}
C_\mathrm{FOV}=\frac{2.24\pi\times10^{22} \times10^{0.4(m_\odot-H_\mathrm{R})}}{p_\mathrm{R}},
\label{eq:C}
\end{eqnarray}

\noindent
where $m_\odot$ = $-27.1$ is the $R_\mathrm{C}$ magnitude of the Sun \citep{Drilling2000}, and $p_\mathrm{R}$ is the geometric albedo in the $R_\mathrm{C}$ band.
Assuming $p_\mathrm{R}$ = 0.04, which is often used for cometary grains, we obtained $C_\mathrm{FOV}$ = (2.3 $\pm$ 0.5) $\times$ $10^{10}$ m$^2$.

The grain mass can be derived as $M_\mathrm{FOV}=\frac{4}{3}C_\mathrm{FOV}~a_\mathrm{d}\rho_\mathrm{d}$, where $a_\mathrm{d}$ and $\rho_\mathrm{d}$ are the grain radius and mass density, respectively. From the synchrone analysis (see Figure 3), the long extended structure extended upper right (along the negative velocity vector, $-v$), which is consistent with dust particles of $a_\mathrm{d}\gtrsim$1 mm but inconsistent with $a_\mathrm{d}\lesssim$100 \micron.  Assuming $\rho_\mathrm{d}$ = 10$^3$ kg m$^{-3}$,
we obtained $M_\mathrm{FOV}$ = 3 $\times$ 10$^{10}$ kg (when $a_\mathrm{d}$ = 1 mm) or $M_\mathrm{FOV}$ = 3 $\times$ 10$^{11}$ kg (when $a_\mathrm{d}$ = 1 cm).
It is important to note that the derived mass is the lower limit because the dust particles extended beyond the FOV.
Nevertheless, we can use the  $M_\mathrm{FOV}$ value to identify the origin of the long extended structure.
The dust production rate of 17P/Holmes was $\sim$3 kg s$^{-1}$ around its perihelion before the 2007
outburst \citep{Ishiguro2013}. Supposing that a comet loses most of its mass at a constant rate within 2.5 AU,
which corresponds to a duration of about a year, 17P/Holmes was expected to lose about 10$^8$ kg in total during the
last perihelion passage, except for the outburst. The mass is significantly (more than two orders of magnitude) smaller than
the mass of the long extended structure in the KWFC FOV. In contrast, the total mass of the outburst ejecta was derived as 10$^{10}$--10$^{13}$ kg \citep{Montalto2008,Altenhoff2009,Reach2010,Ishiguro2010,Boissier2012}.
The mass of the long extended structure is equivalent to or smaller than the total ejecta mass of the 2007 outburst.
Together with the position angle and morphology we mentioned above, we conclude that the long extended structure
is a part of the dust cloud ejected during the 2007 outburst but is tentatively detected by our observation because of the convergence
effect  (i.e., the neck-line effect). In Section \ref{sec:model}, we further investigate the physical  properties of the long extended
structure, regarding it as a remnant dust ejecta during the 2007 outburst.

\section{The Dynamical Model}
\label{sec:model}
We conducted a model simulation
of dust particles ejected during the 2007 outburst to derive the size and ejection velocity of the 2007 outburst ejecta.
The basic theory is essentially the same as that described in \citet{Ishiguro2014}.
In the model, the motion of dust particles is governed by the ejection velocity ($V$) and the solar radiation pressure
(parameterized by $\beta$, the ratio of the radiation pressure acceleration to solar gravity, which is the same as $\beta$ in Section
\ref{subsubsection:tail}). For spherical compact particles
with a mass density of $\rho_\mathrm{d}$ (kg m$^{-3}$) and a radius of $a_\mathrm{d}$ (m), it is written as $\beta$ = 5.7 $\times$ 10$^{-4}$
$\rho$$^{-1}$$a_\mathrm{d}$$^{-1}$ \citep{Burns1979,Finson1968}. The modeled images were generated by the Monte Carlo approach
assuming the velocity and size distribution. Positions of dust particles at the observed epoch were calculated semi-analytically 
by solving the Kepler's equations rigorously.

\subsection{Size estimate}
\begin{figure}
 \epsscale{1.0}
   \plotone{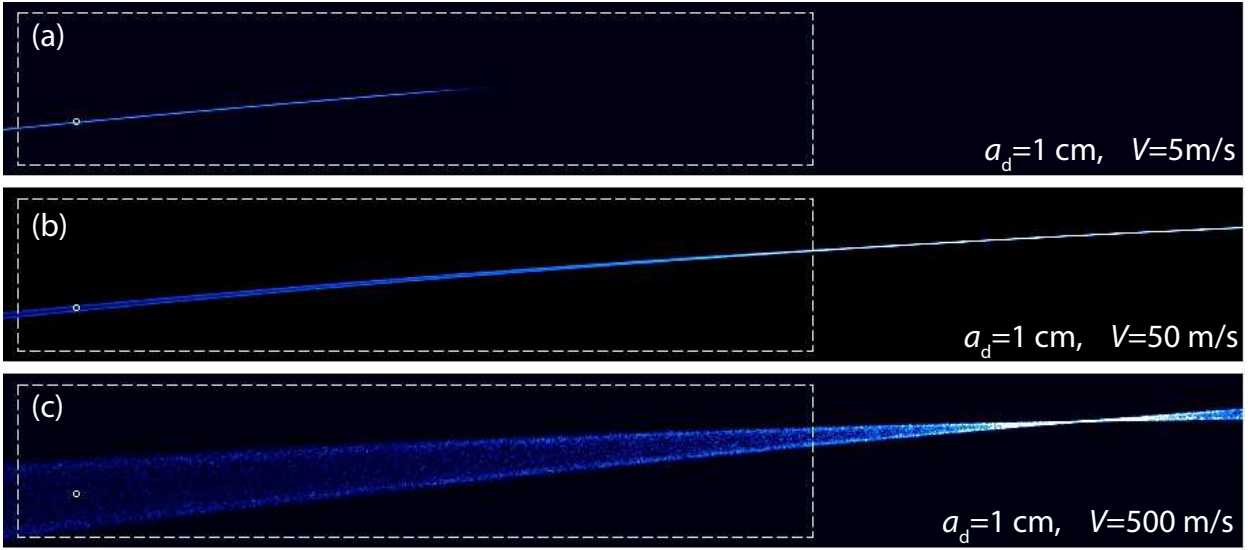}
  \caption{Simulation image of dust particles ejected at the time of the outburst on UT 2007 October 23
  assuming 1 cm grains with different ejection velocities.
  Position of nucleus is indicated by circle with a radius of 60\arcsec\ (equivalent to the observed width).
  The convergent point appears 2.7\arcdeg\ from the nucleus in (b) and (c).
  Dashed rectangle indicates the FOV of the composite image in Figure \ref{fig:f2}. Orientation of the image
  is the same as in Figure \ref{fig:f2}; i.e., north is up, and east is to the left. }
    \label{fig:f6}
\end{figure}

\begin{figure}
 \epsscale{1.0}
   \plotone{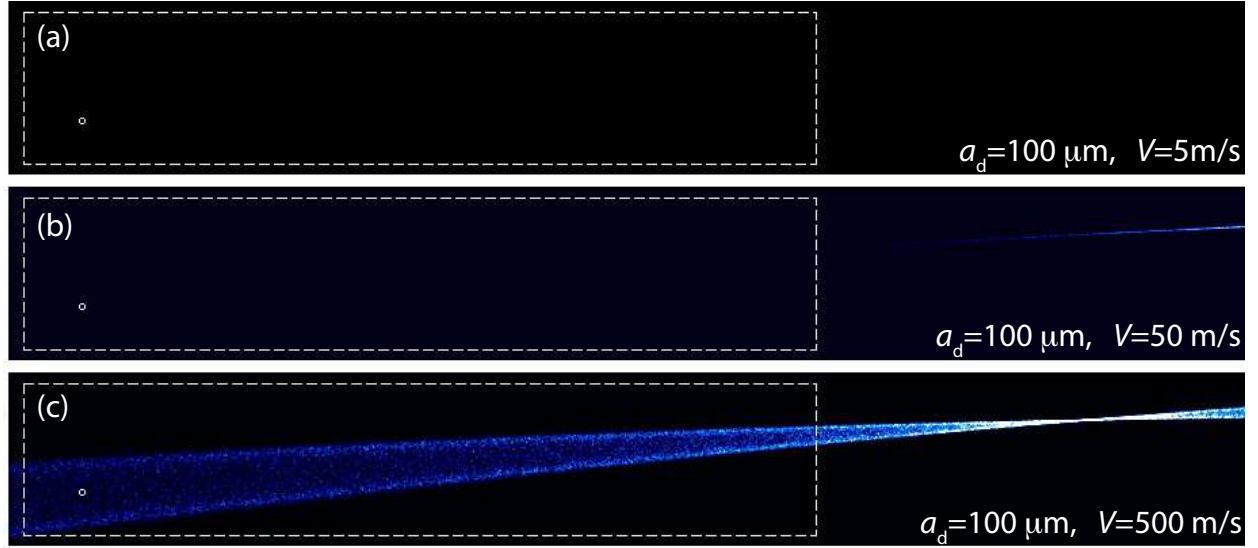}
  \caption{The same as Figure \ref{fig:f6} but for smaller dust particles ($a_\mathrm{d}$ = 100 \micron). Note that the dust cloud is
  beyond the right (i.e., western) edge of the KWFC FOV in (a) and (b) owing to solar radiation pressure.}
    \label{fig:f7}
\end{figure}

As a first step, we considered a simple impulsive dust ejection model in which the dust cloud consists of dust particles with
a uniform size and ejection velocity. Assuming $\rho_\mathrm{d}$ = 10$^3$ kg m$^{-3}$ and isotropic dust ejection (i.e., particles
are ejected equally in all directions), we simulated $a_\mathrm{d}$ = 100 \micron\ particles and $a_\mathrm{d}$ = 1 cm particles
in a possible velocity range of $V$ = 5--500 m s$^{-1}$. The upper limit of $V$ is comparable to the highest dust velocity in the outburst
envelope \citep[i.e., 554 m s$^{-1}$,][]{Lin2009}, whereas the lower limit is close to the escape velocity ($V_\mathrm{es}=1.5$ m s$^{-1}$) from the
nucleus with $R_\mathrm{N}$=2080 m \citep{Stevenson2014} and a bulk density of 1000 kg m$^{-3}$. Figures \ref{fig:f6} and \ref{fig:f7} show the
resultant simulated images. In Figure \ref{fig:f6}, the width and length of the neck-line structure increase as $V$ increases.
The convergent point appears about 2.7\arcdeg\ rightward of the nucleus, showing the strongest intensity enhancement (i.e., neck-line
structure). These models qualitatively reproduce the observed narrowing and brightening from east to west. A visual
comparison yielded an order-of-magnitude estimate of the ejection velocity of $V\approx$ 50 m s$^{-1}$ for 1 cm particles.
For 100 \micron\ particles with low $V<$50 m s$^{-1}$, the dust cloud was blown off westward beyond the FOV,
because such small dust particles are susceptible to solar radiation pressure and are strongly accelerated toward the negative velocity vector
(i.e., western direction) (Figure \ref{fig:f7} (a)--(b)). The increase in ejection velocity would enlarge the dust cloud in every direction, having
the cloud appear in the FOV of the KWFC. However, we noticed that 100 \micron\ is too small to be detected in the KWFC FOV. If we increase $V$ to be
observable in the FOV, it should have a width much wider than what we observed (e.g., Figure \ref{fig:f7} (c)).
To summarize the uniform size model, we applied constraints of $a_\mathrm{d}\gtrsim$ 100 \micron\ and $V\approx$ 50 m s$^{-1}$.

\subsection{Model with arbitrary size and velocity}
Second, we employed a more realistic model having a power-law size distribution
and velocity distribution for the neck-line structure. We assumed the number of dust particles was

\begin{equation}
N(a_\mathrm{d})~da_\mathrm{d} = N_0 \left(\frac{a_\mathrm{d}}{a_{0}}\right)^{q}~da_\mathrm{d}
 \label{eq:size}
\end{equation}

\noindent in the size range $a_\mathrm{min}$ $\le$ $a_\mathrm{d}$ $\le$ $a_\mathrm{max}$,
where $a_\mathrm{min}$ and $a_\mathrm{max}$ are the minimum and maximum particle sizes detectable in the KWFC FOV, respectively.
Further, $a_0$ is the reference size of dust particles (we set $a_0$ = 1 cm and $a_\mathrm{min}$ = 100 \micron\, following the result of
the uniform size model above). We use the following function for the  ejection terminal velocity of dust particles:

\begin{equation}
V = V_0 \left(\frac{a_\mathrm{d}}{a_{0}}\right)^{u} v,
 \label{eq:velocity}
\end{equation}

\noindent where $v$ is a random variable having an average of 1 and a standard deviation $\sigma_v$
\citep[see, e.g.,][]{Ishiguro2014}. We set $\sigma_v$ = 0.1, which was the value obtained from a similar outburst at P/2010 V1. Note that
$v$ makes a minor contribution to the spatial distribution of dust particles when $\sigma_v\ll1$.  Now we have five variables
($N_0$, $a_\mathrm{max}$, $q$, $V_0$, and $u$). Among them,
$N_0$ can be determined by scaling the simulation intensity to the observed one once the other four parameters are fixed.
We thus created a number of  simulation images to find the best-fit parameter set assuming $a_\mathrm{max}=1$ cm, 10 cm, and 1 m
in  $-4.5\leq q \leq -3.0$ at the interval of $\Delta q=0.1$, $1\leq V_0 \leq 70$ m s$^{-1}$ at the interval of $\Delta V_0=1$ m s$^{-1}$,
and $0.1\leq u \leq 0.9$ at the interval of $\Delta u=0.1$, respectively.

The simulation revealed several trends. The width of the neck-line structure increases as $V_0$ increases.
Further, $q$ is also sensitive to the width. The width increases when $q$ is small, because small particles with higher velocity
are efficient scatterers for smaller $q$. We compared the width and intensity distribution with respect to the distance
from the nucleus with those of the model simulation. We found the leftmost data point
does not match any model probably because the data was contaminated by an unconsidered error source (such as the remnant
of background stars or imperfect flat-fielding), and thus ignored it. We obtained the best-fit parameters via $\chi^2$  test,
$q$ = $-$3.4 $\pm$ 0.1, $V_0$ = 50 $\pm$ 10 m s$^{-1}$, and $u$ = 0.3 $\pm$ 0.1.  We appended the errors of these parameters
when simulation results matched the observed results to an accuracy of the measurement.
In the parameter range above, $a_\mathrm{max}$ is less determined (although we got the constraint of $a_\mathrm{max}>1$ cm), because such large particles are supposed to stay near the comet nucleus while such signature from biggest particles was obscured by the bright dust tail and coma. 


\section{DISCUSSION}
\label{sec:discussion}

\subsection{Velocity}
\begin{figure}
 \epsscale{0.8}
   \plotone{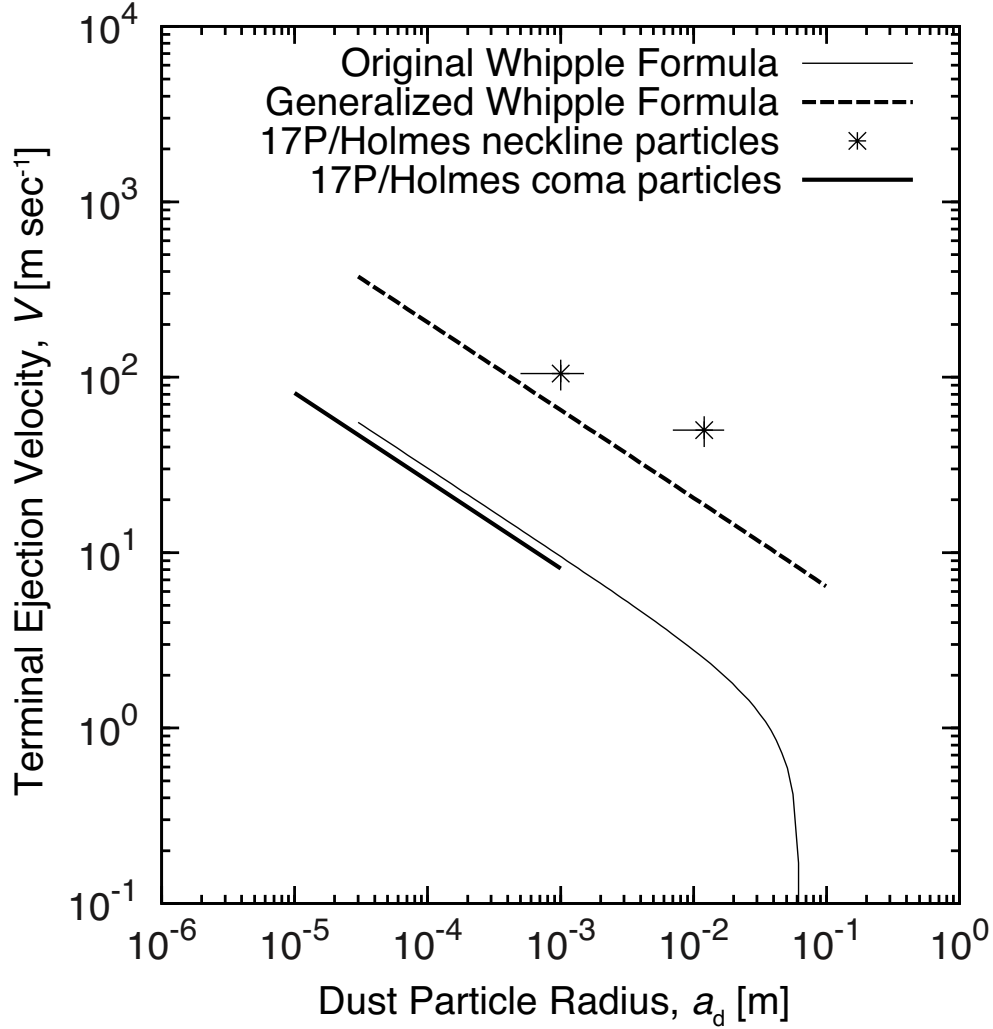}
  \caption{Models of terminal ejection velocity of dust grains ($V$) from 17P/Holmes as a function of particle radius ($a_\mathrm{d}$).
  Ejection velocities of dust grains in the neck-line structure and coma (determined by the sunward extent) are shown for comparison.
  Original and generalized Whipple models with our measurement are plotted assuming the mass densities of the nucleus and dust particles are
  $\rho_\mathrm{N}$ = $\rho_\mathrm{d}$ = 10$^3$ kg m$^{-3}$. Regarding the generalized model, we consider the water production rate
  $Q_\mathrm{H_2O}$ = 1 $\times$ 10$^{30}$ mol s$^{-1}$ and the gas velocity $\bar{v_\mathrm{g}}$ = 550 m s$^{-1}$.}
    \label{fig:Vej}
\end{figure}

An unexpected result is the high ejection velocity for 1 mm--1 cm particles in the neck-line structure. Before this observation, we
predicted an ejection velocity of several meters per second for various reasons. First, we predicted the low ejection velocity
as an analog of cometary dust trails. They consist of 1 mm--1 cm particles with an ejection velocity of several meters per
second, which is marginally larger than that of the escape velocities from kilometer-sized bodies \citep{Sykes1992,Ishiguro2002}.
In addition, the velocity of the smallest (probably 0.1--1 \micron) particles for the 2007 outburst was estimated to be 554 m s$^{-1}$ \citep{Lin2009}.
If we extrapolate  the velocity of 1 cm particles using the inverse square law of the grain size, we would have 2--6 m s$^{-1}$.
Figure \ref{fig:Vej} shows the velocity of the dust particles with respect to the grain size. For comparison, we show
the velocity predicted using the classical model developed by \citet{Whipple1951}. The model has been widely applied
to characterize the velocity--size law. The velocity in the classical model is found to be in good agreement with that of fresh
dust particles in the coma (see Section \ref{subsubsection:tail}) but more than one order of magnitude smaller than those
of dust particles in the neck-line structure. The discrepancy may suggest that dust ejection in the 17P/Holmes outburst
differed from ejection due to normal sublimation by solar heating.

To compensate for the large discrepancy in the ejection velocity, we reviewed the formula in the Whipple model.
In the generalized formula of the Whipple theory, it is written as \citep{Ryabova2013}

\begin{equation}
V = \sqrt{\frac{k_\mathrm{drag}\Delta M \bar{v_\mathrm{g}}}{2\pi R_\mathrm{N}}\frac{A}{m}-\frac{2GM_\mathrm{N}}{R_\mathrm{N}}},
 \label{eq:vej}
\end{equation}

\noindent
where $k_\mathrm{drag}$ is a drag coefficient, usually assumed to be $k_\mathrm{drag}$ = 26/9, and
$m$ and $A$ are the mass and cross-sectional area, respectively, of the dust particles.
For spherical particles, they are written as $m$ = $4/3\pi \rho_\mathrm{d} a_\mathrm{d}^3$ and $A$=$\pi a_\mathrm{d}^2$.
$M_\mathrm{N}$ is the mass of a nucleus having a radius $R_\mathrm{N}$. Assuming a spherical body with a mass density
$\rho_\mathrm{N}$, it is written as $M_\mathrm{N}$ = $4\pi \rho_\mathrm{N} R_\mathrm{N}^3/3$.
Note that the second term in the square root makes a minor contribution when $V$ is significantly larger than the escape velocity $V_\mathrm{es}$,
where $V_\mathrm{es}$ is 1.5 m s$^{-1}$ for 17P/Holmes ($R_\mathrm{N}$ = 2080 m,
and $\rho_\mathrm{N}$ = 10$^3$ kg m$^{-3}$ are assumed). $\Delta M$ is the mass loss rate of the gas. Assuming that dust particles are
accelerated by water molecules, we can write $\Delta M=\mu_\mathrm{H_2O}~Q_\mathrm{H_2O}$. $\bar{v_\mathrm{g}}$ is the velocity of the water vapor outflow.

In this model, there are several unknown parameters.
We adopted $\bar{v_\mathrm{g}}=$ 550 m s$^{-1}$, which corresponds to the maximum velocity of dust particles in the outburst
envelope \citep{Lin2009} and is also  equivalent to the assumed gas velocity in \citet{Dello Russo2008}.
The water production rate, $Q_\mathrm{H_2O}$, was determined by several authors; the obtained values include
$Q_\mathrm{H_2O}$ = (1.2--1.4) $\times$ 10$^{30}$ mol s$^{-1}$ on UT 2007 October 25--27  \citep{Combi2007},
$Q_\mathrm{H_2O}$ = 4.5 $\times$ 10$^{29}$ mol s$^{-1}$ on UT 2007 October 27.6 \citep{Dello Russo2008}, and
$Q_\mathrm{H_2O}$ = 5 $\times$ 10$^{29}$ mol s$^{-1}$ on UT 2007 November 01.2  \citep{Schleicher2009}.
We adopted $Q_\mathrm{H_2O}$ = 1 $\times$ 10$^{30}$ mol s$^{-1}$.  Assuming the mass densities of the dust particles $\rho_\mathrm{d}$ = 10$^3$ kg m$^{-3}$,
we obtained the size--velocity law (thick dashed line in Figure \ref{fig:Vej}). Although there are uncertainties in $Q_\mathrm{H_2O}$ and $\bar{v_\mathrm{g}}$,
the model velocity coincides with the observed velocity to within a factor of $<$2. We guess that this trivial difference can be explained by
the simplifications in the model. For example, the term 2$\pi$ in the denominator of Eq. \ref{eq:vej} was obtained assuming
hemispherical dust emission. It can be $<$2$\pi$ in a realistic case when the dust particles were ejected from limited active areas on the surface,
increasing the ejection velocity \citep[also described in][]{Hughes2000}. In addition, the gas velocity would be increased via adiabatic expansion into space.
These effects may result in a dust velocity higher than that in the generalized Whipple model.
We thus conclude that a sudden sublimation of a large amount of water ice ($\approx$10$^{30}$ mol s$^{-1}$) would be responsible for
the high ejection velocity of the remnant dust debris in our observed image.

\subsection{Mass and Total Kinetic Energy of the Outburst Grains}

The total mass of the outburst ejecta has been derived by many researchers using different techniques.
It differs substantially depending on the author, i.e., 10$^{10}$--10$^{14}$ kg \citep{Montalto2008,Altenhoff2009,Reach2010,Ishiguro2010,Boissier2012,Li2011,Ishiguro2013}.
As discussed in many papers, there is an intrinsic problem in the determination of the mass of cometary dust because the scattering
cross section is dominated by the smallest particles, whereas the mass of the largest particles in the differential
size frequency distribution has a power index of $-3<q<-4$. The dust cloud of 17P/Holmes may be
no exception. \citet{Zubko2011} studied the polarimetric property of 17P/Holmes outburst ejecta and
found that the size distribution has the power index $q\sim-3.5$.

From our dynamical model, we found that the total cross section of the dust grains in the KWFC FOV
is about 5\%--10\% of the total dust cloud in the size range of $a_\mathrm{d}$ = 100 \micron--1 cm.
Considering the size distribution with an index of $q$ = $-$3.4 $\pm$ 0.1 in the size range
$a_\mathrm{d}$ = 100\micron--1cm, which were obtained by the dynamical model in Section \ref{sec:model},
we obtained a mass of $M_\mathrm{>100\micron}$=(3.6--7.2) $\times$ 10$^{11}$ kg for the neck-line particles (i.e. big remnant particles).
If we integrate down to submicron particles \citep[i.e., $a_\mathrm{min}$ = 0.6 \micron, ][]{Zubko2011}, we derive a total mass
of $M_\mathrm{TOT}$=(3.8--7.7) $\times$ 10$^{11}$ kg for the 2007 outburst ejecta, which is almost the same as $M_\mathrm{>100\micron}$
because largest grains make up most of the mass. The mass also shows good consistency with several
previous studies such as a radio continuum observation by \citet{Boissier2012} and is consistent with the upper limit in \citet{Li2011}.
The kinetic energy of large dust particles ($a_\mathrm{d}$ = 100\micron--1cm) is estimated to be $E_\mathrm{>100\micron}$=(7--14)$\times$10$^{14}$ J.
Similarly, if we integrate down to 0.6 \micron-sized particles, we got $E_\mathrm{TOT}$=(1--6)$\times$10$^{15}$ J.
The energy per unit mass is less than 20\% of the energy released during the crystallization of amorphous water ice.
As already described in \citet{Li2011}, our results may imply that the outburst can be caused by the crystallization of buried amorphous ice.

It is interesting to notice that a large amount of big dust particles was injected into the
interplanetary space by a single outburst event and survived for $>$7 years without melting or disintegrating the structure
and observed as the neck--like structure.
The cometary dust particles will probably disperse in the interplanetary field via planetary perturbations
 and constitute a portion of zodiacal cloud \citep{Vaubaillon2004}. There is a long-standing question
about the origin of zodiacal dust cloud because it erodes by Poynting-Robertson effect and mutual collision
among particles.  We estimated the contribution of cometary dust particles via 17P-like outbursts.
It is not clear how often such big cometary outbursts occurred. We assumed the frequency of 0.1--1 per century
because the comet exhibited similar (but slightly weaker) outburst in the 19th century \citep{Reach2010}. Multiplying
the frequency by $M_\mathrm{TOT}$, we obtained a crude estimate of 12--240 kg sec$^{-1}$. Although there should be a
large uncertainty in the rate,  ejecta by 17P-like outburst would account for a considerable fraction (1--24\%) of the required
mass to sustain the zodiacal cloud \citep[i.e. 10$^{4}$ kg sec$^{-1}$, ][]{Mann2006}. Therefore, we speculate that some fraction of zodiacal dust might be generated by 17P-like outbursts 
and remain in the interplanetary space for a long time to be observable as zodiacal light.

\section{SUMMARY}
\label{sec:summary}
We observed 17P/Holmes in 2014 September using a wide-field imaging camera, the
KWFC attached to the Kiso 105 cm Schmidt Telescope. We found that:

\begin{enumerate}
\item{17P/Holmes consisted of two components: a near-nuclear fresh dust tail and a long extended structure. }
\item{The long structure extended along the position angle of 274.6\arcdeg $\pm$ 0.1\arcdeg\ on UT 2014 September 22 and
showed westward brightening and narrowing, confirming that it was composed of dust grains
ejected during the 2007 outburst}.
\item{The FWHM and intensity are in the range of 30--70\arcsec\ and (1.5--3.1) $\times$ 10$^{-8}$ W m$^{-2}$ sr$^{-1}$ \micron$^{-1}$
in the $R_\mathrm{C}$ band}
\item{The typical size of the particles is 1 mm--1 cm on the basis of a comparison with a dynamical model.}
\item{The ejection speed is around 50 m s$^{-1}$ or even more, which is faster than the values of submicron grains extrapolated using a simple $-$1/2 law.
	We conjecture that the high velocity would result from the sudden sublimation of the icy component.}
\item{The total mass and kinetic energy of the 2007 outburst ejecta were (4--8)$\times$10$^{11}$ kg and  and (1--6)$\times$10$^{15}$ J.
These are consistent with but more accurate than previous studies.}
 \end{enumerate}

\newpage
\vspace{2cm}

This observation was conducted until one day before a terrible volcanic eruption occurred at Mt. Ontake, which is located
about 15 km from the observatory. We offer our sincere sympathy to the victims and hope for a rapid recovery for both
the people and areas affected by the disaster.

\vspace{0.5cm}

Acknowledgments\\
We acknowledge the staff at Kiso Observatory for their kind support in a homey atmosphere.
Thanks to their maintenance of the telescope and instruments, we could make the observation
without any breakdown. This research was supported by a National Research Foundation of Korea (NRF)
grant funded by the Korean government (MEST) (No. 2012R1A4A1028713).
DK is supported by Optical \& Near-Infrared Astronomy Inter-University Cooperation Program, the MEXT of Japan.


\newpage

\newpage

\newpage

\newpage

\newpage

\newpage

\end{document}